\documentclass[prl,twocolumn,aps,showpacs,10pt]{revtex4}
\usepackage{graphicx}% complex graphics
\usepackage{bm}% bold math
\usepackage{amssymb} %math symbols
\usepackage{color}

\begin{document}

\title{Dipole-induced localized plasmon modes and resonant surface plasmon scattering}

\author{E. G. Mishchenko}

\affiliation{Department of Physics and Astronomy, University of Utah, Salt Lake City, UT 84112, USA}

\begin{abstract}
A metal film supports the continuum of propagating surface plasmon waves. The interaction of these waves with a dipole (nanoparticle) positioned some distance from the surface of the film can produce well defined localized plasmon modes whose frequency nonetheless resides inside the continuum. This leads to the resonant enhancement of scattering of surface plasmons off the dipole. The maximum of scattering is found to occur when the distance from the dipole to the surface of the film is equal to one half of the film thickness. The possibility of controllable plasmon scattering could be advantageous for the field of nanoplasmonics.
\end{abstract}
%\pacs{73.22.Pr,73.20.At,73.22.Dj}

\maketitle

{\it Introduction}. Plasmonics is a relatively new field \cite{SAM,HAA,ZZX} whose objective is to explore and utilize collective charge excitations that exist at the interfaces of metals and dielectrics -- surface plasmons, or plasmon-polaritons. Plasmonics applications hold  promise to facilitate optical sub-wavelength resolution \cite{KV}. This is the consequence of two factors: surface plasmons propagate with velocities well below the speed of light, and they can be excited optically in the appropriate geometrical configurations \cite{OKR}. This means that optically excited plasmons are characterized by much smaller wavelengths than those of the light used to excite them, potentially pushing resolution  beyond the diffraction limit.
In the limit of short wavelengths the dispersion condition for surface plasmons requires that the dielectric functions at both sides of the interface add up to zero: e.g. $\varepsilon +1=0$ in the case of a flat metal surface exposed to vacuum \cite{R}. This yields the frequency of surface plasmons, $\omega_0/\sqrt{2}$, significantly reduced compared with the frequency of bulk plasmons, $\omega_0$.
In addition to such propagating plane-wave plasmons related charge oscillations occur in metallic nanoparticles, the standing waves \cite{KZ}, often called localized surface plasmon resonances. For a spherical particle the dipole mode ($l=1$) has the lowest frequency, $\omega_0/\sqrt{3}$. Localized plasmon modes of nanoparticles are of acute interest to plasmonics as the corresponding resonant near-field enhancement is conductive to the improvements in the imaging resolution.

It is well known that localized electron states are often formed in the vicinity of a lattice defect \cite{AM}. In the case of surface plasmons the mere presence of a surface imperfection does not produce a bound plasmon excitation, though it results in a scattering of surface plasmons \cite{ZSM,BDE,LSS,SDZ,O,SHD,KHL,Lee,BHL} that could be crucial to their controlled propagation in photonic devices. Nonetheless, a combination of the metal surface and a localized dipole (nanoparticle, molecule, etc.) produces configuration resonances \cite{KXB}. In the present paper we consider plasmon spectrum of a system consisting of a dipole above a metal film of finite thickness. Configuration resonances, which could be viewed as localized plasmon modes acquire intrinsic attenuation originating from the overlap with the propagating modes, but under the most interesting conditions this attenuation can be controlled via the film thickness with the excitations being well defined. Below we first elucidate how their  properties depend on the geometry of the system before exploring how the corresponding resonances can be revealed in the scattering of propagating surface plasmons off the nanoparticle.
\begin{figure}[h]
\resizebox{.38\textwidth}{!}{\includegraphics{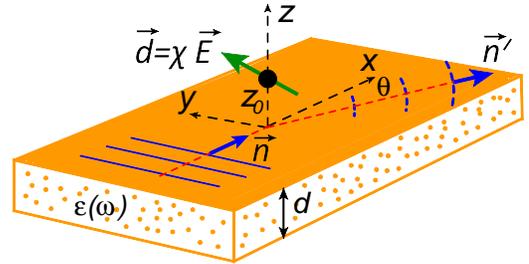}}
\caption{A nanoobject positioned a distance $z_0$ above a metal film of thickness $d$. The dipole acquires the dipole moment in response to the applied electric field. The system depicted here supports a continuum of propagating modes, Eq.~(\ref{propagating_spectrum}), as well as two localized plasmons, Eq.~(\ref{localized_energy}). Since the localized solutions overlap with the continuum they acquire a finite width. Conversely, the localized modes show up as resonances in the scattering of propagating plasmons, indicated schematically with (blue) arrows.}
\label{fig1}
\end{figure}

Consider a point dipole positioned at ${\bf r}={\bf r}_0$ and characterized by the dynamic polarizability $\chi_\omega$, such that the induced dipole moment ${\bf d}=  \chi_\omega {\bf E}$ is proportional to the electric field ${\bf E}$ oscillating with frequency $\omega$. For the time being we assume that the dipole is surrounded by some arbitrary configuration of dielectric and conducting media. Neglecting retardation effects is legitimate if the characteristic length scales of the problem are much smaller than $c/\omega$, in which case the electric properties of the system are described by the Poisson's equation for the scalar potential:
\begin{equation}
\label{poisson}
\nabla ( \varepsilon({\bf r}) \nabla \varphi) = -4\pi \rho({\bf r}),
\end{equation}
where $\rho({\bf r})$ is the density of extraneous charges. In the case of a point dipole $\rho({\bf r})=-{\bf d} \cdot \nabla \delta ({\bf r}-{\bf r}_0)$. Using the Green's function $G({\bf r},{\bf r}')$ of Eq.~(\ref{poisson}) the potential created by the dipole is expressed as
\begin{equation}
\label{Poiss_solution}
\varphi({\bf r}) = -4\pi \nabla' G({\bf r},{\bf r}_0) \cdot {\bf d},
\end{equation}
where $\nabla'$ indicates the derivative taken with respect to the second argument of the Green's function.
The latter, in the vicinity of the dipole (where the dielectric constant $\kappa_0$ does not depend on the position), consists of two parts:
\begin{equation}
G({\bf r},{\bf r}') =-\frac{1}{4\pi \kappa_0|{\bf r}-{\bf r}'|} +{\cal G}({\bf r},{\bf r}'),
\end{equation}
where the first term describes the point source and the second (non-singular) contribution stands for the field of the induced charges distributed throughout the rest of the system. The dipole moment can now be expressed as ${ d}_i= 4\pi \chi_\omega \partial_i \partial_j' {\cal G}({\bf r}_0,{\bf r}_0)d_j$. The non-trivial solutions exist when
\begin{equation}
\label{localized_dispersion}
\det{\Bigl(\delta_{ij}- 4\pi \chi_\omega  \partial_i \partial_j' {\cal G}({\bf r}_0,{\bf r}_0)\Bigr)}=0.
\end{equation}

\begin{figure}[h]
\resizebox{.40\textwidth}{!}{\includegraphics{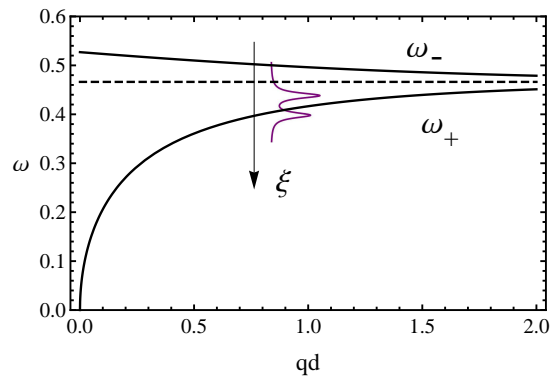}}
\caption{Dispersion (in units of $\omega_0$) of plasmons, Eq.~(\ref{propagating_spectrum}), in a silver ($\varepsilon_b=3.6$, $\omega_0=1.4 \times 10^{16}$ s$^{-1}$) film. Positions of localized states, Eq.~(\ref{dispesion_localized}), are shown schematically. The variable $\xi=(1+\varepsilon(\omega))/(1-\varepsilon(\omega))$ characterizes the deviation of the frequency from the limiting value $\omega_0/\sqrt{\varepsilon_b+1}$, shown by the dashed line.}
\label{fig_new1}
\end{figure}

{\it Dipole-induced modes in a metal film}. We now apply this equation to the situation of a dipole surrounded by vacuum ($\kappa_0=1$) and positioned at a point ${\bf r}_0 = (0,0,z_0)$ above a metal film of finite thickness $d$, see Fig.~\ref{fig1}. The film is infinite in the $xy$-plane and occupies space between $-d<z<0$. It is characterized by the dielectric function $\varepsilon(\omega)=\varepsilon_b-\omega_0^2/\omega^2$, where $\omega_0$ is the (bulk) plasma frequency \cite{ELM}. Finding the corresponding Green's function is straightforward in the Fourier representation with respect to the in-plane coordinates $\bm \rho =(x,y)$, which is done by  taking into account the appropriate boundary conditions. As a result, in the region above the film,
$z,z'>0$:
\begin{equation}
\label{green_second}
{\cal G}(q,z,z')=\frac{1}{2q}\frac{(\varepsilon^2-1)(e^{qd}-e^{-q d}) e^{-q(z+z')}}{(\varepsilon+1)^2e^{qd}-
(\varepsilon-1)^2e^{-qd}}.
\end{equation}
The denominator of this Green's function has two poles,
\begin{equation}
\label{propagating_spectrum}
\omega_\pm^2(q)=\omega_0^2 \frac{1\mp e^{-qd}}{\varepsilon_b+1\mp (\varepsilon_b-1)e^{-qd}},
\end{equation}
corresponding to the well-known surface plasmon modes \cite{R}, characterized by the symmetric (upper sign) and anti-symmetric (lower sign) distribution of the induced surface charges, see Fig.~\ref{fig_new1}.
Taking the inverse Fourier transform of the Green's function (\ref{green_second}) and substituting it into the characteristic equation for the localized modes
(\ref{localized_dispersion}) we obtain that there are two solutions ($n=1,2$),
\begin{equation}
\label{dispesion_localized}
\frac{n}{2\alpha}=  {\cal D}_\omega, ~~~~{\cal D}_\omega \equiv z_0^3 \int\limits_0^\infty dq ~ q^3 {\cal G}(q;z_0,z_0),
\end{equation}
corresponding to the dipole moment polarized along the $z$-axis ($n=1$) and within the $xy$-plane ($n=2$). Here we have introduced the dimensionless dipole strength, $\alpha = \chi_\omega/z_0^3$. The modes determined by Eq.~(\ref{dispesion_localized}) have finite width even in the absence of any electron scattering. This happens because the localized plasmon energy overlaps with the frequency of the propagating modes (\ref{propagating_spectrum}). The imaginary part then arises from the pole contributions to the integral in Eq.~(\ref{dispesion_localized}). As we are going to see, the resulting attenuation is weak, however, and the localized plasmons are well defined provided some geometrical conditions are satisfied.

It is convenient to proceed by introducing the dimensionless variable $2qz_0=t$ in Eq.~(\ref{dispesion_localized}):
\begin{eqnarray}
\label{dimensionless_dispersion}
{\cal D}_\omega = \frac{\xi}{16} ~\text{P}  \int\limits_0^\infty dte^{-t} t^2 \frac{1-e^{-t(d/z_0)}}{\xi^2-e^{-t(d/z_0)}}\nonumber\\
+i\frac{\pi z^3_0}{2d} \widetilde{q}^2 e^{-2\widetilde q z_0} \sinh{\left(\widetilde q d\right)}.
\end{eqnarray}
 where the dimensionless frequency variable $\xi=(1+\varepsilon(\omega))/(1-\varepsilon(\omega)$ is defined in such a way that the positive values $\xi>0$ overlap with the symmetric surface plasmons, cf.~Eq.~(\ref{propagating_spectrum}), while the negative values fall inside the anti-symmetric domain.
The imaginary part in Eq.~(\ref{dimensionless_dispersion}) originates from the momenta $\widetilde q$ determined by $|\xi|=\exp{(-\widetilde qd)}$. The meaning of $\widetilde q$ is the momentum of the propagating surface plasmons into which the localized state decays.

Let us emphasize that $\alpha$ in Eq.~(\ref{dispesion_localized}) is in general a function of frequency and can be either positive or negative. Large vales of $\alpha$ are possible near an internal dipole resonance.  Under less favorable conditions, i.e. away from the resonances, however, the effective coupling constant is expected to be small, $\alpha \ll 1$. Here we consider in more detail this weak coupling limit, which imposes more stringent conditions on the existence of the localized modes.

In order to satisfy Eq.~(\ref{dispesion_localized}) for small values of the coupling constant $\alpha$ the real part of the function ${\cal D}_\omega$ has to be large, something that in any case requires that $\xi\ll 1$. The principal value integral in Eq.~(\ref{dimensionless_dispersion}) is calculated in this limit to be,
\begin{eqnarray}
\label{function_D}
{\cal D}_\omega' &=& \frac{\pi z_0^3~\text{sign}\xi}{4d^3|\xi|^{1-2z_0/d}} \Bigl( {\cal C}\ln^2{|\xi|}  -\pi (1+{\cal C}^2)\ln{|\xi |} \nonumber\\ && +\frac{\pi^2}{2} {\cal C}(1+{\cal C}^2)\Bigr), ~~~~~~~{\cal C}=\cot\left(\frac{\pi z_0}{d} \right).
\end{eqnarray}
This expression gives an excellent approximation for all values of $d>z_0$ with the exception of a very narrow (logarithmically) range where $d$ is very close to $z_0$.
More quantitatively, the condition $|\xi|^{2-2z_0/d} \ll 1$ has to be satisfied. As we are going to see shortly this is indeed the situation where the most interesting results are manifested.

(i) {\it Thick films}. When the thickness of the film is much greater than the distance from its surface to the dipole the number ${\cal C}$ is large. This forces $\xi$ to be close to $\alpha/4n$, which is indeed its exact limit when $d\to \infty$. For finite but still large $d\gg z_0$ the following formula gives an excellent numerical agreement with the exact solution for all $d/z_0 \gtrsim 5$: $\xi=\frac{1}{4n}\alpha [1-\frac{1}{2}(z_0/d)^3 \ln^3{(\frac{4n}{|\alpha|})}]$, as long as $\alpha$ is not too large ($\lesssim 0.5$).
This yields the spectrum of localized plasmons (for $\varepsilon_b=1$),
\begin{equation}
\label{localized_energy}
\omega^2=\frac{\omega^2_0}{2}\left(1-\frac{\alpha}{4n} +\frac{ \alpha  z_0^3}{8nd^3} \ln^3\frac{4 n}{|\alpha|} -\frac{i\pi |\alpha | z_0^3}{2nd^3} \ln^2\frac{4 n}{|\alpha|} \right).
\end{equation}
In the limit $d=\infty$ Eq.~(\ref{localized_energy}) recovers the result of Ref.~\onlinecite{KXB}.
%Note that  the logarithmic correction to the real part here strictly speaking has a rather limited meaning as the imaginary part is of the same (or greater) magnitude.
Finally, in addition to the intrinsic dissipation into propagating plasmons, explicitly given by Eq.~(\ref{localized_energy}), there might be additional contribution from the imaginary part of $\alpha$ not considered here.
\begin{figure}[h]
\resizebox{.45\textwidth}{!}{\includegraphics{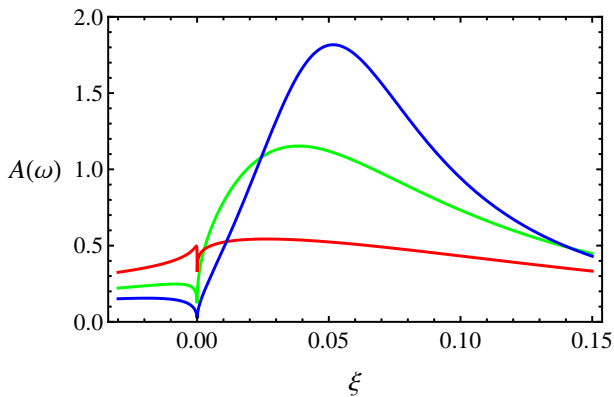}}
\caption{The spectral weight $A(\omega)=\text{Im}~{(1-2\alpha {\cal D}_\omega)^{-1}}$ of the $z$-polarized ($n=1$) localized plasmon mode (for $\alpha =0.3$) as a function of the dimensionless ``distance'' $\xi$ from the frequency $\omega_0/\sqrt{\varepsilon_b+1}$, listed in the order of decrease of the maximum height: $d/z_0=$2.5 (blue), 2.0 (green), 1.5 (red).}
\label{fig_new2}
\end{figure}

(ii) {\it Thin films}. With decreasing the ratio $d/z_0$ the function (\ref{function_D}) decreases (for a given coupling $\alpha$) \cite{KK}. When $d=2z_0$ a critical situation arises as the power-law behavior is replaced with the logarithmic dependence. Here ${\cal C}=0$ and $\xi=\exp{(-\frac{16 n}{\pi^2\alpha})}$. For yet smaller values of $d/z_0<1$ the function ${\cal D}_\omega$ becomes non-monotonic and drops sharply at $\xi \to 0$. This would indicate the existence of two solutions (per given value of $n$) of Eq.~(\ref{dispesion_localized}). However, the increase in the imaginary part of ${\cal D}_\omega$ renders distinguishing such solutions impossible. It is straightforward to obtain a quantitative condition ensuring the sharpness of the localized plasmon resonances. Such condition is inferred  from Eq.~(\ref{localized_energy}) by requiring that the real part of $\xi$ exceeds its imaginary part:
\begin{equation}
\label{sharpness_condition}
\frac{d}{z_0} \gtrsim \left(2\pi \ln^2\frac{4}{|\alpha|}\right)^{1/3}.
\end{equation}
Fig.~\ref{fig_new2} illustrates the spectral weight of the localized modes for different values of the film thickness.
The spectral weight as a function of frequency can be found from considering the dipole's response to a local ac field ${\cal E}_z$ (e.g. for probing the $z$-polarized state). With the help of Eq.~(\ref{Poiss_solution}) the resulting dipole moment can be found to be, $d_z=\chi_\omega {\cal E}_z/(1-4\pi \chi \partial_z^2 {\cal G})$. The energy dissipation rate is thus given by, $\langle \dot{d}_z {\cal E}_z \rangle =\frac{\omega}{2}\chi_\omega|{\cal E}_z|^2\text{Im}~{(1-2\alpha {\cal D}_\omega)^{-1}}$.
For films of large thicknesses where the condition (\ref{sharpness_condition}) is well satisfied the peaks are sharp and symmetric, but with decreasing $d$ display clear Fano-type asymmetry.
Let us emphasize, however, that even for films of smaller thickness the localized plasmon modes, while broadened, could still affect various response functions significantly. One of such phenomena will be considered now.

{\it Surface plasmon scattering}. The very situation that gives rise to the finite lifetime of the localized modes is advantageous for their observation. Since the localized modes overlap with the propagating surface plasmons (\ref{propagating_spectrum})
the scattering of the latter off the dipole should exhibit {\it resonances} at the corresponding plasmon energies \cite{EB}.
The scattering amplitude of a surface plasmon incident on the dipole with the in-plane momentum ${\bf n}k$, see Fig.~\ref{fig1},
is determined with the help of the asymptotic form of the scalar potential at large distances from the scatterer ($\rho \to \infty, ~ z>0$),
\begin{equation}
\label{incident_scattered}
\varphi({\bm \rho},z)=  e^{\pm i k{\bf n}\cdot{\bm \rho}-kz}+ \frac{f_\pm(\theta)}{\sqrt{\rho} } e^{\pm i k {\bf n'}\cdot{\bm \rho}-kz},
\end{equation}
for the symmetric (upper sign) and anti-symmetric modes, respectively.
The unit vector ${\bf n'}$ indicates the direction of the scattered wave's propagation, $\cos\theta ={\bf n}\cdot {\bf n}'$.
The $z$-dependence of the scalar potential outside the film in Eq.~(\ref{incident_scattered}) follows from the Laplace's equation. The time dependence $\propto e^{-i\omega t}$ is dropped for brevity.
Note that for the anti-symmetric plasmon the direction of propagation is {\it opposite} to the direction of its momentum ($-{\bf n}$), since its group velocity is negative, cf. Eq.~(\ref{propagating_spectrum}).
The incident field, given by the first term in Eq.~(\ref{incident_scattered}), which we henceforth denote by $\varphi^{(0)}$, must be added to the previously considered solution (\ref{Poiss_solution}) of the Poisson's equation, yielding in the Fourier representation,
\begin{equation}
\label{phi_new}
\varphi({\bf q},z) = \varphi^{(0)}({\bf q},z)+4\pi \left(  i{\bf q}\cdot {\bf d}_\parallel -d_z \partial_{z_0}\right) G(q;z,z_0).
\end{equation}
The induced dipole moment ${\bf d}=- \chi_\omega \nabla  \Phi ({\bf r}_0)$ is given by the field of all extraneous charges (excluding the dipole's own field) $\Phi({\bf r})$. This ``outside'' field can again be easily separated from the total field in Eq.~(\ref{phi_new}) if only the ``induced'' term (\ref{green_second}) in the Green's function is retained:
\begin{eqnarray}
%\label{h_out}
\Phi({\bm \rho},z) &=& \varphi^{(0)}({\bm \rho},z)-4\pi  \chi_\omega
\int \frac{d^2 q}{(2\pi)^2}  e^{i{\bf q} {\bm \rho}} \nonumber\\ &\times& \Bigl[i{\bf q} \cdot\nabla_{\bm \rho} \Phi ({\bf r}_0)  -
  \partial_z \Phi({\bf r}_0) \partial_{z} \Bigr]{\cal G}(q;z,z_0), ~~~
\end{eqnarray}
where the derivatives of the potential in the right-hand side are taken at the origin $(0,0,z_0)$. By calculating now the gradient of both sides of this expression  we obtain,
\begin{equation}
\nabla_{\bm \rho} \Phi  ({\bf r}_0)=\frac{\displaystyle \nabla_{\bm \rho} \varphi^{(0)} ({\bf r}_0) }{ 1-\alpha {\cal D}_\omega},~~~~~
 \partial_z \Phi({\bf r}_0)=\frac{\displaystyle  \partial_z \varphi^{(0)}({\bf r}_0) }{ 1-2\alpha {\cal D}_\omega}.
\end{equation}
We can now write for the scattered field defined as, $\Phi_{sc}=\Phi-\varphi_0$, the following expression,
\begin{eqnarray}
\label{h_out}
\Phi_{sc}({\bm \rho},z) &=&2  \alpha z_0^3  k e^{-k z_0} \int\limits_0^\infty dq q^2 {\cal G}(q;z,z_0) \nonumber \\ & \times &
\Bigl[\frac{J_0(q\rho)}{1-2\alpha  {\cal D}_\omega}\pm  i \cos\theta \frac{J_1(q\rho)}{1-\alpha {\cal D}_\omega} \Bigr].~~~~~~
\end{eqnarray}
The scattered field $\Phi_{sc}$ does not contain the field of the dipole itself. The latter, however, is short-range and thus does not contribute to the asymptotic behavior described by Eq.~(\ref{incident_scattered}). The expression (\ref{h_out}) is non-perturbative in the coupling constant $\alpha$ and as such describes multiple scattering events.

\begin{figure}[h]
\resizebox{.45\textwidth}{!}{\includegraphics{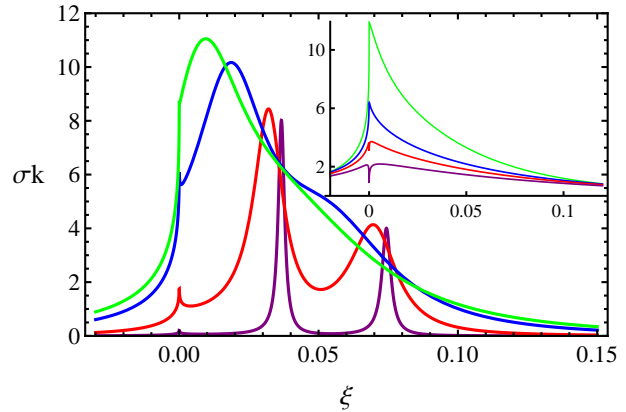}}
\caption{Total dimensionless scattering cross section $\sigma k$ as a function of the dimensionless plasmon frequency $\xi=1-2\omega^2/\omega_0^2$,  for $\alpha=0.3$. Main frame shows four different curves, from bottom to top: $d/z_0=$10 (purple), 5 (red), 3 (blue), 2.5 (green). Inset: from top to bottom curves, $d/z_0=$ 2, 1.7, 1.6, 1.5.}
\label{fig2}
\end{figure}

To find the scattering amplitude it is sufficient to calculate only the asymptotics of the integral in Eq.~(\ref{h_out}). This is most easily performed by replacing Bessel functions with the Hankel functions according to, $J_n =(H_n^{(1)}+H_n^{(2)})/2$, and followed by the  rotation of the integration contour in the complex plane so that it runs along the positive or negative imaginary semi-axis (for the terms containing Hankel functions of the first and second order, respectively). The asymptotic value of the integral is then determined by the pole singularities, see Eq.~(\ref{green_second}), that expectedly originate from the contributions of propagating plasmons. As a result, for the two propagating modes the scattering amplitudes amount to,
\begin{eqnarray}
\label{amplitude}
f_\pm (\theta) =\mp  \sqrt{\frac{8}{\pi k}}e^{\pm i\pi/4}  \left(\frac{\alpha D_\omega''}{1-2\alpha {\cal D}_\omega}+\frac{\alpha D_\omega'' \cos\theta }{1-\alpha  {\cal D}_\omega} \right).
\end{eqnarray}
Interestingly, the scattering amplitude explicitly depends only on the geometry of the system, but not on the metal properties (i.e. the concentration of carriers). The two poles in Eq.~(\ref{amplitude}) describe the resonant scattering by the localized plasmon modes discussed above. Since the propagating surface plasmons have both the in-plane and out-of-plane components of the electric field both modes appear in the scattering amplitude. Fig.~\ref{fig2} illustrates the behavior of the total scattering cross-section $\sigma=\int d\theta |f(\theta)|^2$ (identical for both symmetric and anti-symmetric modes) for various values of the ratio $d/z_0$.

In agreement with the condition (\ref{sharpness_condition}) the resonant peaks in the cross section are sharp for large values of $d/z_0$. The heights of the peaks are $8/k$ and $4/k$ for the scattering from localized plasmons polarized in- and out-of-plane, respectively, independent of the strength of the coupling $\alpha$. This is evident from the fact that the imaginary part of ${\cal D}_\omega$ dominates the denominators in Eq.~(\ref{amplitude}) near the resonances. With decreasing the ratio $d/z_0$ the peaks broaden and merge together while shifting towards $\xi=0$. In addition a sharp asymmetric feature emerges at $\xi=0$  related to the singularity in the density of states  in the propagating plasmon spectrum. The cross section reaches its maximum in the critical regime of $d/z_0=2$ and then  drops sharply for the yet thinner films. In addition, another peculiar feature, the suppression of scattering at $\xi=0$ emerges eventually. It can be traced to the overlap of the localized state with the propagating plasmons described by the imaginary part of ${\cal D}_\omega$, see Eq.~(\ref{dimensionless_dispersion}). At $\xi \to 0$, when the momentum $\widetilde q$ becomes infinity the overlap has a distinctly different behavior for $d>2z_0$ (where it diverges) and for $d<2z_0$, where it is suppressed.

{\it Radiative losses}. So far we have neglected radiative losses from oscillating motion of the dipole and the electrons in the metal film. Such losses are small by the same condition already mentioned in the introduction that allows to neglect retardation effects. The radiative reaction field acting on the dipole can be estimated from the classic formula ${\bf E}_{rad}= 2\dddot{ \bf d}/{3c^3}$. Requiring that this field is negligible compared with the (electrostatic) field of the induced charges, $4\pi \partial_i \partial'_j {\cal G}(z_0,z_0) d_j$. From Eqs.~(\ref{dimensionless_dispersion}) and (\ref{function_D})  we get $\xi \ll (c/\omega z_0)^3$.

{\it Experimental realization}. When the dipole describes a metallic nanoparticle with the dielectric function $ \widetilde \varepsilon$ and radius $a$ its polarizability is $\alpha=\frac{4\pi}{3}(\frac{a}{z_0})^3 (\widetilde \varepsilon-1)/(\widetilde \varepsilon +2)$. To realize the $\alpha \sim 1$ situation the most straightforward way is to produce the nanoparticle from such a metallic material that ensures that its first internal resonance, determined by $\widetilde \varepsilon(\omega) = -2$, is close to the extended plasmon frequency of the metal film, $\varepsilon(\omega) = - 1$. This seems to be feasible for Ag films, where $\varepsilon_b=3.6$ and $\omega_0=75~256$ cm$^{-1}$  \cite{FW,RC}, and Pd particles, where $\widetilde \varepsilon =1-\widetilde \omega_0^2/\omega^2$, with $\widetilde \omega_0=60~495$ cm$^{-1}$  \cite{YS,RC}. This gives for the frequency of propagating plasmons in Ag, $\omega_0/\sqrt{\varepsilon_d+1}= 35~088$ cm$^{-1}$, very close to the first resonance of Pd spheres, at $\widetilde \omega_0/\sqrt{3} =34~927$ cm$^{-1}$. As a result for $a=10$ nm and $z_0 = 100$ nm, we obtain $\alpha\approx -0.45$.

{\it Summary and discussion}. In an analogy to the appearance of localized electron states near a crystal defect, a point dipole positioned above a metal film surface can produce localized plasmon modes. Such modes split off from the frequency $\omega_0/\sqrt{\varepsilon_b+1}$ by the amount proportional to the dimensionless dipole strength, see Eq.~(\ref{localized_energy}). One of the modes is characterized by the dipole oscillating in-plane, while the other one is polarized perpendicularly to the metal surface. Both modes overlap with the continuum of the propagating plasmons and therefore acquire a finite linewidth. The modes are sharp as long as the film thickness $d$ is greater than the distance to the dipole, see Eq.~(\ref{sharpness_condition}). While this aspect of the coupling could be viewed as a detriment, it opens the possibility of probing the localized states via plasmon scattering. In thick films the scattering is dominated by the two narrow peaks whose heights are rather universal and do not depend on the coupling strength or the geometry (the ratio $d/z_0$). With decreasing $d$ the scattering cross section increases and broadens into an asymmetric resonance. The maximum cross section is reached for $d=2z_0$. With the further decrease in the film thickness the scattering drops sharply.
Our results should be of interest to nanoplasmonics, both for the direct purpose of increasing near fields with localized plasmon resonances, as well as via the controlled guiding of propagating plasmons. In the latter case the desired degree of control could be achieved by changing the position of the nano-dipole.

Useful discussions with M.~Raikh % and O.~Starykh
are gratefully
acknowledged. The work was supported by the Department of Energy,
Office of Basic Energy Sciences, Grant No.~DE-FG02-06ER46313.


\begin{thebibliography}{50}

\bibitem{HAA} H.A. Atwater, Sci. Am. {\bf 296}, 56 (2007).

\bibitem{SAM} S.A. Maier, {\it Plasmonics: Fundamentals and Applications}
(Springer, New York, 2007).

\bibitem{ZZX} J. Zhang, L. Zhang, and W. Xu, J. Phys. D: Appl. Phys. {\bf 45}, 113001 (2012).

\bibitem{KV} U. Kreibig and M. Vollmer, {\it Optical Properties of Metal Clusters} (Springer, Berlin, 1995).

\bibitem{OKR} A. Otto, Z. Phys. {\bf 216}, 398 (1968); E. Kretschmann and H. Raether, Z. Naturforsch. A {\bf 23}, 2135 (1968).

\bibitem{R} R.H. Ritchie, Phys. Rev {\bf 106}, 874 (1957).

\bibitem{KZ} U. Kreibig and P. Zacharias, Z. Phys. {\bf 231}, 128 (1970).

\bibitem{AM} Ch. Kittel, {\it Quantum Theory of Solids} (Wiley, 1987).

\bibitem{ZSM} A.V. Zayats, I.I. Smolyaninov, and A.A. Maradudin, Phys. Rep. {\bf 408}, 131 (2005).

\bibitem{BDE} W.L. Barnes, A. Dereux, and T.W. Ebbesen, Nature (London) {\bf 424}, 824 (2003).

\bibitem{LSS} Z. Liu, J.M. Steele, W. Srituravanich, Y. Pikus, C. Sun, and X. Zhang, Nano Lett. {\bf 5}, 1726 (2005).

\bibitem{SDZ} I.I. Smolyaninov, C.C. Davis, and A.V. Zayats, New J. Phys. {\bf 7}, 175 (2005).

\bibitem{O} E. Ozbay, Science {\bf 311}, 189 (2006).

\bibitem{SHD} I.I. Smolyaninov, Y.-J. Hung, and C.C. Davis, Phys. Rev. B {\bf 76}, 205424 (2007).

\bibitem{KHL} H. Kim, J. Hahn, and B. Lee, Opt. Express {\bf 16}, 3049 (2008).

\bibitem{Lee} J.Y. Lee et al., Nature (London) {\bf 460}, 498 (2009).

\bibitem{BHL} B. Baumeier, F. Huerkamp, T.A. Leskova, and A.A. Maradudin, Phys. Rev. A {\bf 84}, 013810 (2011).

\bibitem{KXB} O. Keller, M. Xiao, and S. Bozhevolnyi, Surf. Sci. {\bf 280}, 217 (1993).

\bibitem{ELM} In silver this Drude model expression is a fairly good approximation of the optical response, while in gold it is acceptable away from the interband transitions (at $\sim 470$ nm and $\sim 330$ nm), P.G. Etchegoin, E.C. Le Ru, and M. Meyer, J. Chem. Phys. {\bf 125}, 164705 (2006).

\bibitem{KK} In the limit of $d\ll z_0$, not analysed in the present paper, the situation of a dipole above the two-dimensional electron gas is realized, see V.A. Kochelap and S.M. Kukhtaruk, J. Appl. Phys. {\bf 109}, 114318 (2011); S.M. Kukhtaruk, Ukr. J. Phys. {\bf 55}, 916 (2010).

\bibitem{EB} Non-resonant scattering by a dipole was previously considered for $d=\infty$ by A.B. Evlyukhin and S.I. Bozhevolnyi, Phys Rev. B {\bf 71}, 134304 (2005); T. S$\o$ndergaard and S.I. Bozhevolnyi, Phys. Rev. B {\bf 69}, 045422 (2004).

\bibitem{FW} G.W. Ford and W.H. Weber, Phys. Rep. {\bf 113}, 195 (1984).

\bibitem{RC} R. Rojas and F. Claro, J. Chem. Phys. {\bf 98}, 998 (1992).

\bibitem{YS}  A.Y-C. Yu and W.E. Spicer, Phys. Rev. {\bf 169}, 497 (1968).

\end{thebibliography}
\end{document}